\documentclass[12pt]{article}
\usepackage[applemac]{inputenc}
\usepackage{amssymb,amsmath}
\usepackage{bm}
\title{Calculation on the Lifetime of Polarized Muons in Flight}
\author{Zhi-Qiang Shi$^1$, Guang-Jiong Ni$^{2,3}$\\[0.3cm]
{\it \small $^1$Department of Physics, Shaanxi Normal University, Xi'an 710062, China}\\
    email:{\tt \small zqshi@snnu.edu.cn}\\[0.3cm]
{\it \small $^2$Department of Physics, Fudan University, Shanghai 200433, China}\\
{\it \small $^3$Department of Physics, Portland State University, Portland OR 97207, USA}\\
    email:{\tt \small pdx01018@pdx.edu}}

\begin{document}
\maketitle \vskip 1cm
\begin{abstract}
The fermion lifetime was usually calculated in terms of spin states. However, after comparing spin states with helicity
states and chirality states, it is pointed out that a spin state is helicity degenerate, and so it can not be used for
discussion of the dependence of lifetime on the polarization of an initial-state fermion. Using helicity states, we
calculate the lifetime of polarized muons. The result shows that the lifetime of right-handed polarized muons is always
greater than that of left-handed polarized muons with the same speed in flight.\\[0.3cm]
\end{abstract}

\section{Introduction}

In a previous Letter$^{[1]}$ we have proposed the lifetime asymmetry of left-right handed polarized fermions. Based on
parity violation in the standard model, there exist only left-handed (LH) chirality states in charged weak currents.
Therefore, in weak interaction not only the final-state fermions, but also the initial-state fermions should reveal the
feature of longitudinal polarization. It results in that the right-handed (RH) polarized fermion lifetime
$\tau_{_{Rh}}$ is different from the LH polarized fermion lifetime $\tau_{_{Lh}}$, i.e.
\begin{equation}\label{eq:1}
    \tau_{_{Rh}}=\frac{\tau}{1-\beta}\quad \hbox{and}\quad
    \tau_{_{Lh}}=\frac{\tau}{1+\beta},
\end{equation}
where $\tau$ is the average lifetime, $\tau=\tau_{_0}/\sqrt{1-\beta^2}$, in which $\beta$ is the velocity of the
fermions and $\tau_{_0}$ is the lifetime in its rest frame. $\tau_{_{Rh}}$ and $\tau_{_{Lh}}$ transform to each other
under a space inversion. It is shown that the lifetime of RH polarized fermions is always greater than that of LH
polarized fermions in flight with the same speed in any inertial frame. In this Letter, the lifetime asymmetry will be
further investigated. We will discuss the relation among spin states, helicity states and chirality states before
calculating the lifetime of polarized muons concretely.

\section{Spin State, Helicity State and Chirality State}
\subsection{Spin State}
There exist three kinds of spinor wave functions, i.e., spin states, helicity states and chirality states. The spin
states are the plane wave solutions of Dirac equation, and in momentum representation for a given four-momentum $p$ and
mass $m$, the positive energy solution and the negative energy solution are respectively
\begin{equation}\label{eq:2}
  u_s(p)=\sqrt{\frac{E_p+m}{2E_p}}\left(\!\begin{array}{c}\varphi_s\\
  \frac{\displaystyle \bm\sigma\cdot\bm p}{\displaystyle E_p+m}\varphi_s\;
  \end{array}\!\right)\;,
\end{equation}
\begin{equation}\label{eq:3}
  v_s(p)=\sqrt{\frac{E_p+m}{2E_p}}\left(\!\begin{array}{c}\frac{\displaystyle \bm \sigma\cdot\bm p}
  {\displaystyle E_p+m}\varphi_s\\\varphi_s
  \end{array}\!\right)\;,
\end{equation}
where $s=1,\; 2$ and $\varphi_s$ are Pauli spin wave functions. The state with $s=1$ is spin up while the state with
$s=2$ is spin down. They are eigenstates of operator, $\frac{\omega(p)\cdot e}{m}$, with eigenvalues $\pm 1$, namely
\begin{equation}\label{eq:4}
  \frac{\omega(p)\cdot e}{m}\;u_s(p)=\left\{\begin{array}{lr}u_s(p),\quad&(s=1)\\
  -u_s(p).\quad&(s=2)\end{array}\right.
\end{equation}
The $\omega(p)$ is the Pauli-Lubanski covariant spin vector and $e$ is the four-polarization vector in the form
\begin{equation}\label{eq:5}
  e_{\alpha}=\left\{\begin{array}{ll}\bm e^0+\frac{\displaystyle\bm p\;(\bm p\cdot\bm e^0)}
  {\displaystyle m(E_p+m)}\;,\quad &(\alpha=1,2,3)\\
  i\;\frac{\displaystyle\bm p\cdot\bm e^0}{\displaystyle m}\;,\quad &(\alpha=4)
  \end{array}\right.
\end{equation}
which is normalized $(e^2=1)$, orthogonal to $p\;(e\cdot p=0)$. In the rest frame $e$ reduces to $e^0=(\bm
e^0,0)=(0,0,1,0)$. In application it is frequently necessary to evaluate spin sums in the form
\begin{equation}\label{eq:6}
  \begin{array}{l}
  P_1(p)=\rho_+\;\Lambda_+(p),\quad P_2(p)=\rho_-\;\Lambda_+(p),\\
  P_3(p)=\rho_+\;\Lambda_-(p),\quad P_4(p)=\rho_-\;\Lambda_-(p).
  \end{array}
\end{equation}
The $\Lambda_+(p)$ and $\Lambda_-(p)$ are the positive energy projection operator and the negative energy projection
operator,
\begin{equation}\label{eq:7}
  \Lambda_+(p)=\frac{-i\;\gamma\cdot p+m}{2E_p},\quad\Lambda_-(p)=\frac{i\;\gamma\cdot p+m}{2E_p},
\end{equation}
respectively and $\rho_{\pm}$ are spin projection operators:
\begin{equation}\label{eq:8}
  \rho_{\pm}=\frac{1}{2}(1\pm i\;\gamma_5\;\gamma\cdot e).
\end{equation}
The plus sign refers to $s=1$ and the minus sign to $s=2$. We have eigenvalue equations
\begin{eqnarray}\label{eq:9}
  \rho_+\;u_1(p)&=&u_1(p),\quad\rho_-\;u_2(p)=u_2(p)\;,\nonumber\\
  \rho_-\;u_1(p)&=&\rho_+\;u_2(p)=0\;.
\end{eqnarray}
One sees that the operator $P_1(p)$ project out the positive energy state with spin up and $P_2(p)$ the positive energy
state with spin down, whereas the operator $P_3(p)$ the negative energy state with spin down and $P_4(p)$ the negative
energy state with spin up in its rest frame.

\subsection{Helicity State}
A helicity state is the eigenstate of the helicity of fermions. If spinor $\varphi$ is taken as the eigenstate of the
spin component along the direction of its motion,
\begin{equation}\label{eq:10}
  \frac{\bm \sigma\cdot
  \bm p}{|\bm p|}\;\varphi_{_h}=h\;\varphi_{_h},\quad h=\pm 1
\end{equation}
then the helicity states read
\begin{equation}\label{eq:11}
  u_{_h}(p)=\sqrt{\frac{E_p+m}{2E_p}}\left(\!\begin{array}{c}\varphi_{_h}\\
  \frac{\displaystyle h|\bm p|}{\displaystyle E_p+m}\varphi_{_h}
  \end{array}\right)\;.
\end{equation}
The state with $h=+1$ is the RH helicity state while the state with $h=-1$ is the LH helicity state. The projection
operators of the helicity states are$^{[2]}$
\begin{equation}\label{eq:12}
  \rho_{_h}=\frac{1}{2}\left(1\pm \frac{\displaystyle \bm \Sigma\cdot \bm p}{\displaystyle |\bm p|}\;\right)\;.
\end{equation}
We can see that the helicity state is entirely different from the spin state, the helicity eigenvalue is not Lorentz
invariant and the projection operator of helicity state is essentially a two-component operator. From
Eqs.~(\ref{eq:2}), (\ref{eq:3}) and (\ref{eq:4}) we can find out that a spin state with the same $s$ but different
values of $h$ is helicity degenerate and so it can not uniquely describe the helicity of fermions. The eigenvalue
equations (\ref{eq:9}) depend only on the quantum number $s$ and remain valid for an arbitrary helicity state ($h=-1$
or $+1$). It implies that the spin projection operators $\rho_{\pm}$ can only project out the states which in its rest
frame have spin $s=1$ and $2$, respectively. Strictly speaking, only in the rest frame can the four-polarization vector
$e$ and the eigenvalue equation (4) be most unambiguously defined$^{[3]}$. Taking the simplest case of $\bm p: \bm
p=p_z$, which does not lose the universality of problem, we have the LH helicity state $u_{_{Lh}}$ and the RH helicity
state $u_{_{Rh}}$, respectively
\begin{eqnarray}
  u_{_{Lh}}(p)&=&\sqrt{\frac{E_p+m}{2E_p}}\left(\!\begin{array}{c}\varphi_2\\
  \frac{\displaystyle {-|\bm p|\;\varphi_2}}{\displaystyle {E_p+m}}\end{array}\!\right),
  \label{eq:13}\\
  u_{_{Rh}}(p)&=&\sqrt{\frac{E_p+m}{2E_p}}\left(\!\begin{array}{c}\varphi_1\\
  \frac{\displaystyle {|\bm p|\;\varphi_1}}{\displaystyle {E_p+m}}
  \end{array}\!\right).\label{eq:14}
\end{eqnarray}
Even so comparing Eq.~(\ref{eq:8}) with Eq.~(\ref{eq:12}) one can also see that the projection operator of spin state
is different from that of helicity state though the spin state and the helicity state are formally identical when $\bm
p=p_z$. Hence we reach a conclusion that the polarization of fermions must be described by the helicity states which
are closely related to directly observable quantity experimentally.

\subsection{Chirality State}
The chirality states are the eigenstates of chirality operator $\gamma_5$. The LH chirality state and the RH chirality
state are defined as, respectively
\begin{equation}\label{eq:15}
  u_{_{LS}}(p)=\!\frac{1}{2}(1+\gamma_5)u_s(p),\quad
  u_{_{RS}}(p)=\!\frac{1}{2}(1-\gamma_5)u_s(p).
\end{equation}
In general, chirality states are different from helicity states. Only if $m=0$ (for example neutrinos) or $E\gg m$ (in
the ultrarelativistic limit) the fermions satisfy Weyl equation$^{[1,4]}$, the spinor $\varphi_s$ must then be taken to
be eigenstates of helicity operator $h$ and the polarization is always in the direction of motion$^{[5]}$. In other
words, for $m=0$ the helicity states, the chirality states and spin states are identical, i.e.
\begin{eqnarray}
  u^W_{Lh}(p)&=&u^W_{L2}(p)=u^W_2(p)=\frac{1}{\sqrt{2}}\left(\begin{array}{c}
  \varphi_2\\-\varphi_2\end{array}\right),\label{eq:16}\\
  u^W_{Rh}(p)&=&u^W_{R1}(p)=u^W_1(p)=\frac{1}{\sqrt{2}}\left(\begin{array}{c}
  \varphi_1\\\varphi_1\end{array}\right).\label{eq:17}
\end{eqnarray}
The superscript $W$ refers to it being a solution of Weyl equation.

The helicity states $u_{_{Lh}}$ and $u_{_{Rh}}$ in Eqs.~(\ref{eq:13}) and (\ref{eq:14}) can be expanded as linear
combination of chirality states, respectively
\begin{eqnarray}
    u_{_{Lh}}(p)&=&\frac{1}{2}(1+\gamma_5)\;u_{_{Lh}}(p)+\frac{1}{2}(1-\gamma_5)\;u_{_{Lh}}(p)\nonumber\\
    &=&C_{LL}\;u^0_{_{L2}}+C_{LR}\;u^0_{_{R2}},\label{eq:18}\\
    u_{_{Rh}}(p)&=&C_{RL}\;u^0_{_{L1}}+C_{RR}\;u^0_{_{R1}},\label{eq:19}
\end{eqnarray}
where $u^0_{_{LS}}$ and $u^0_{_{RS}}$ are chirality states in the rest frame,
\begin{equation}\label{eq:20}
  u^0_{_{LS}}=\frac{1}{2}\left(\!\begin{array}{c}\varphi_s\\-\varphi_s
  \end{array}\!\right),\quad
  u^0_{_{RS}}=\frac{1}{2}\left(\!\begin{array}{c}\varphi_s\\\varphi_s
  \end{array}\!\right).
\end{equation}
It can be seen from Eqs.~(\ref{eq:18}) and (\ref{eq:19}) that the decompositions of helicity states do not possess
Lorentz invariance and the $\bm p$ dependence has been shifted to the coefficients $C_{LL}$, $C_{LR}$, $C_{RL}$ and
$C_{RR}$ as given by
\begin{eqnarray}
    C_{LL}=C_{RR}&=&\frac{1}{\sqrt{2E_p}}(\sqrt{E_p+m}+\sqrt{E_p-m})\nonumber\\
    &=&\sqrt{1+\beta},\label{eq:21}\\
    C_{RL}=C_{LR}&=&\frac{1}{\sqrt{2E_p}}(\sqrt{E_p+m}-\sqrt{E_p-m})\nonumber\\
    &=&\sqrt{1-\beta}.\label{eq:22}
\end{eqnarray}
It is obvious from Eqs.~(\ref{eq:18}) and (\ref{eq:19}) that in a LH helicity state $u_{_{Lh}}(p)$ the coefficient
$C_{LL}$ is the amplitude of LH chirality state $u^0_{_{L2}}$ and the $C_{LR}$ that of RH chirality state
$u^0_{_{R2}}$; while in a RH helicity state $u_{_{Rh}}(p)$ the $C_{RL}$ that of LH chirality state $u^0_{_{L1}}$ and
the $C_{RR}$ that of RH chirality state $u^0_{_{R1}}$ in its rest frame.

\section{The lifetime of polarized muons}

Now let us consider a $\mu$ decay process
\begin{equation}\label{eq:23}
    \mu^-\rightarrow e^-+\nu_\mu+\bar{\nu}_e\;.
\end{equation}
The lowest order decay rate or lifetime $\tau$ for muon decays, based on the perturbation theory of weak interactions,
is given by
\begin{equation}\label{eq:24}
  \tau^{-1}=\frac{1}{(2\pi)^5}\!\int\!d^3 q\;d^3 k\;d^3 k'\;\delta^4(p-q-k-k')\;M^2.
\end{equation}

\subsection{The lifetime of unpolarized muons}
If the muons are unpolarized and if we do not observe the polarization of final-state fermions, then the transition
matrix element is given by averaging over the muon spin and summing over all final fermion spins:
\begin{eqnarray}\label{eq:25}
  M^2&=&\frac{G^2}{2}\;\frac{1}{2}\sum_{s,s',r,r'=1}^2\left[\;\overline{u}_{s'}(q)\;
  \gamma_\lambda\;(1+\gamma_5)\;v_{r'}(k')\;\right]^2\nonumber\\
  &&\times\left[\;\overline{u}_r(k)\;\gamma_\lambda\;(1+\gamma_5)\;u_s(p)\;\right]^2.
\end{eqnarray}
where $p$, $q$, $k$ and $k'$ are 4-momenta, while $s$, $s'$, $r$ and $r'$ are spin indices for $\mu$, $e$, $\nu_\mu$
and $\bar{\nu}_e$, respectively. For the convenience of discussion below, in Eq.~(\ref{eq:25}) we set
\begin{equation}\label{eq:26}
  I=\frac{1}{2}\sum_{s=1}^2\sum_{r=1}^2\left[\;\overline{u}_r(k)\;\gamma_\lambda\;
  (1+\gamma_5)\;u_s(p)\;\right]^2,
\end{equation}
which is related to the muons. By means of Eqs.~(\ref{eq:6}) and (\ref{eq:8}), the evaluations of spin sums are reduced
to the calculation of projection operators:
\begin{eqnarray}
  \sum_{s=1}^2u_s(p)\overline{u}_s(p)&=&\sum_{s=1}^2P_s(p)=\Lambda_+(p),\label{eq:27}\\
  \sum_{s=1}^2v_s(p)\overline{v}_s(p)&=&-\sum_{s=1}^2P_s(p)=-\Lambda_-(p).\label{eq:28}
\end{eqnarray}
One sees that the explicit evaluations of spin projection operators disappear. Applying Eqs.~(\ref{eq:27}),
(\ref{eq:28}) and (\ref{eq:7}) as well as the trace theorems we obtain
\begin{equation}\label{eq:29}
  M^2=\frac{4\;G^2\;(p\cdot k')\;(q\cdot k)}{E_p E_q E_k E_{k'}}.
\end{equation}
Substituting Eq.~(\ref{eq:29}) into Eq.~(\ref{eq:24}), one has
\begin{equation}\label{eq:30}
  \tau^{-1}=\frac{4\;G^2}{(2\pi)^5}\frac{1}{E_p}\int\!\frac{d^3 q}{E_q}\frac{d^3 k}{E_{k}}\frac{d^3
  k'}{E_{k'}}\delta^4(p-q-k-k')\cal F,
\end{equation}
where
\begin{equation}\label{eq:31}
  {\cal F}=(p\cdot k')\;(q\cdot k).
\end{equation}
Obviously decay amplitude $\cal F$ is a Lorentz-invariant matrix element. Therefore we have
\begin{equation}\label{eq:32}
  {\cal F}={\cal F}^0=(p^0\cdot k')(q\cdot k),\quad p^0=(0,0,0,im_\mu)
\end{equation}
where ${\cal F}^0$ is $\cal F$ in the muon rest frame.

The integration to the right of $E_p^{-1}$ in Eq.~(\ref{eq:30}) is a Lorentz scalar, and we see that the lifetime is
proportional to the energy $E_p$ as required by special relativity$^{[6]}$. So that the lifetime is not a Lorentz
scalar.

Starting from Eq.~(\ref{eq:30}) and neglecting electron mass, the muon lifetime in its rest frame is given by
\begin{equation}\label{eq:33}
    \tau^{-1}_0=\frac{G^2m^5_\mu}{192\;\pi^3},
\end{equation}
where $m_\mu$ is muon mass. The $\tau_0$ is the muon lifetime in its rest frame and defined as the muon lifetime in
ordinary tables of particle properties. In an arbitrary frame the muon lifetime is given by
\begin{equation}\label{eq:34}
  \tau^{-1}=\sqrt{1-\beta^2}\;\tau_0^{-1}.
\end{equation}

\subsection{The lifetime of polarized muons expressed by spin states}
For polarized muons the muon spin should not be averaged. In most literatures and textbooks$^{[7,8]}$ the polarized
states of fermions were usually expressed by the spin states (\ref{eq:2}) and (\ref{eq:3}). Instead of
Eqs.~(\ref{eq:26}) and (\ref{eq:27}), we have
\begin{equation}\label{eq:35}
  I_s=\sum_{r=1}^2\left[\overline{u}_r(k)\gamma_\lambda(1+\gamma_5)u_s(p)\right]^2,
\end{equation}
and
\begin{equation}\label{eq:36}
  u_s(p)\overline{u}_s(p)=P_s(p)=\frac{1}{2}(1\pm i\gamma_5\gamma\cdot e)\frac{(-i\gamma\cdot p+m_\mu)}{2E_p},
\end{equation}
respectively. Substituting Eq.~(\ref{eq:36}) into (\ref{eq:35}) and Eq.~(\ref{eq:35}) into (\ref{eq:25}) we obtain
\begin{equation}\label{eq:37}
  M_s^2=\frac{4\;G^2{\cal F}_s}{E_p E_q E_k E_{k'}},
\end{equation}
where
\begin{equation}\label{eq:38}
  {\cal F}_s=(p\cdot k')\;(q\cdot k)\mp m_\mu(e\cdot k')\;(q\cdot k).
\end{equation}
Obviously, decay amplitude ${\cal F}_s$ is also a Lorentz scalar, like $\cal F$. Therefore we have
\begin{equation}\label{eq:39}
  {\cal F}_s={\cal F}^0_s=(p^0\cdot k')\;(q\cdot k)\mp m_\mu(e^0\cdot k')\;(q\cdot k),
\end{equation}
where ${\cal F}^0_s$ is ${\cal F}_s$ in the muon rest frame.

In a similar way to Eq.~(\ref{eq:30}), we obtain
\begin{equation}\label{eq:40}
    \tau^{-1}_s=\frac{4\;G^2}{(2\pi)^5}\frac{1}{E_p}\int\!\frac{d^3 q}{E_q}\frac{d^3
    k}{E_{k}}\frac{d^3k'}{E_{k'}}\delta^4(p-q-k-k'){\cal F}_s.
\end{equation}
One easily verifies that the integration over the second term of decay amplitude ${\cal F}_s$ vanishes. It is easy to
see that the lifetime in the laboratory frame is certainly identical with the result (\ref{eq:34}), i.e.,
$\tau_s=\tau$, which does not exhibit any lifetime asymmetry.

\subsection{The lifetime of polarized muons expressed by helicity states}
As mentioned above, however, this method does not enable us to discuss the dependence of lifetime on the helicity of
muons. Because a spin state is helicity degenerate, the polarization of muons must be described by helicity states. For
LH polarized muons, substituting the spin states in Eq.~(\ref{eq:35}) with the helicity states and considering the
energy projection operator of helicity state is equal to that of spin state, we obtain
\begin{eqnarray}
    I_{Lh}&=&\sum_{h}\left[\overline{u}_h(k)\gamma_\lambda
    (1+\gamma_5)u_{_{Lh}}(p)\right]^2\nonumber\\
    &=&\sum_{r=1}^2\left[\overline{u}_r(k)\gamma_\lambda
    (1+\gamma_5)u_{_{Lh}}(p)\right]^2.\label{eq:41}
\end{eqnarray}
From Eqs.~(\ref{eq:18}), (\ref{eq:21}) and (\ref{eq:15}) we easily find
\begin{equation}\label{eq:42}
  (1+\gamma_5)u_{_{Lh}}(p)=2\sqrt{1+\beta}u^0_{_{L2}}=\sqrt{1+\beta}(1+\gamma_5)u^0_2,
\end{equation}
where $u^0_2$ is the spin state in the muon rest frame. One can see that the chirality-state projection operator
$(1+\gamma_5)$ picks out LH chirality state $u^0_{_{L2}}$ in a LH helicity state, which is factorized into two parts in
the second equation, one is the spin state $u^0_2$ and another is a factor $\sqrt{1+\beta}$ which depends on muon's
helicity. Substituting Eq.~(\ref{eq:42}) into Eq.~(\ref{eq:41}) we have
\begin{equation}\label{eq:43}
    I_{Lh}=(1+\beta)\sum_{r=1}^2\left[\overline{u}_r(k)\gamma_\lambda
    (1+\gamma_5)u^0_2\right]^2
\end{equation}
Comparing Eq.~(\ref{eq:43}) with Eq.~(\ref{eq:35}) and considering Eq.~(\ref{eq:39}) we find out the decay amplitude of
LH polarized muons
\begin{equation}\label{eq:44}
  {\cal F}_{Lh}=(1+\beta){\cal F}_2^0=(1+\beta){\cal F}_2.
\end{equation}

Similarly, for RH polarized muons we obtain
\begin{eqnarray}\label{eq:45}
    I_{Rh}&=&\sum_{h}\left[\overline{u}_h(k)\gamma_\lambda
    (1+\gamma_5)u_{_{Rh}}(p)\right]^2\nonumber\\
    &=&(1-\beta)\sum_{r=1}^2\left[\overline{u}_r(k)\gamma_\lambda
    (1+\gamma_5)u^0_1\right]^2
\end{eqnarray}
and the decay amplitude is
\begin{equation}\label{eq:46}
    {\cal F}_{Rh}=(1-\beta){\cal F}_1^0=(1-\beta){\cal F}_1.
\end{equation}
Obviously, both the polarized muon's decay amplitude ${\cal F}_{Lh}$ and ${\cal F}_{Rh}$ are not Lorentz scalar.
Comparing Eqs.~(\ref{eq:44}) and (\ref{eq:46}) with Eq.~(\ref{eq:38}), respectively and considering Eq.~(\ref{eq:40}),
we find the polarized muon lifetimes which agree with Eq.~(\ref{eq:1}).

\section{Summary and Discussion}

The calculation has established that the lifetime of RH polarized muons is greater than that of LH polarized muons in
flight. Furthermore, this conclusion is also valid for all fermions in the decays under weak interactions. Hence under
the condition of parity violation the lifetime is neither a four-dimensional scalar, nor a scalar under the
three-dimensional space inversion. We emphasize here an important concept that a spin state is helicity degenerate and
the spin projection operators $\rho_{\pm}$ can only project out the spin states, but can not project out the helicity
states. The above calculation shows that the so-called eigenstate of four-dimensional spin operator given by
Eq.~(\ref{eq:4}) is by no means a helicity eigenstate (even we had chosen $\bm p$ vector along z axis) and this is why
the parity violation result, Eq.~(\ref{eq:1}), was overlooked in the past for so long a time even one did not perform
the spin average for muons in the laboratory frame. Therefore, the polarized fermions must be expressed by the helicity
states which may satisfy the ordinary Dirac equation$^{[9]}$ and are relevant to physical interpretation and
experimental tests. In the helicity states $u_{_{Lh}}(p)$ and $u_{_{Rh}}(p)$, Eqs.~(\ref{eq:18}) and (\ref{eq:19}),
since the charged weak current originates from the LH chirality state only,$^{[1]}$  the terms corresponding to
$C_{LR}$ and $C_{RR}$ do not contribute to fermion decay. The lifetime of polarized fermions depends only on the
amplitude $C_{LL}$ or $C_{RL}$, which is the root cause of the lifetime asymmetry. In the rest frame, because
$C_{LL}=C_{RL}=1$, the lifetime of LH polarized fermions is equal to that of RH polarized fermions.

The measurements on muon decay used to be performed in its rest frame. They were realized that muons, formed by forward
decay in flight of pions inside cyclotron, were stopped in a nuclear emulsion, sulphur, carbon, calcium or polyethylene
target. For example, the polarization effects of muon decay were observed using carbon stopping target$^{[10]}$, in
which there is no depolarization of the muons. So far the measurement of the lifetime of polarized muons in flight has
not yet been found in literature. Therefore, one actually lacks direct experimental evidence either to support or to
refute the lifetime asymmetry. We report it here now in the hope that it may stimulate and encourage further
experimental investigations on the question of the lifetime asymmetry in muon decays.

\section*{acknowledgments}

We are grateful to Dr. Valeri V. Dvoeglazov and Dr. LiMing for their many helpful discussions.

\end{document}